# Balancing Bounded Treewidth Circuits

Maurice Jansen[*]     Jayalal Sarma M.N.[*]

### Abstract

Algorithmic tools for graphs of small treewidth are used to address questions in complexity theory. For both arithmetic and Boolean circuits, it is shown that any circuit of size $n^{O(1)}$ and treewidth $O(\log^i n)$ can be simulated by a circuit of width $O(\log^{i+1} n)$ and size $n^c$, where $c = O(1)$, if $i = 0$, and $c = O(\log \log n)$ otherwise. For our main construction, we prove that multiplicatively disjoint arithmetic circuits of size $n^{O(1)}$ and treewidth $k$ can be simulated by bounded fan-in arithmetic formulas of depth $O(k^2 \log n)$. From this we derive the analogous statement for syntactically multilinear arithmetic circuits, which strengthens the central theorem of [MR08]. As another application, we derive that constant width arithmetic circuits of size $n^{O(1)}$ can be balanced to depth $O(\log n)$, provided certain restrictions are made on the use of iterated multiplication. Also from our main construction, we derive that Boolean bounded fan-in circuits of size $n^{O(1)}$ and treewidth $k$ can be simulated by bounded fan-in formulas of depth $O(k^2 \log n)$. This strengthens in the non-uniform setting the known inclusion that $SC^0 \subseteq NC^1$. Finally, we apply our construction to show that REACHABILITY for directed graphs of bounded treewidth is in LogDCFL.

## 1   Introduction

It is well-known that many hard graph theoretical problems become tractable when restricted to graphs of bounded treewidth[1]. If a graph with $n$ nodes has bounded treewidth, there always exists a balanced tree decomposition of depth $O(\log n)$. This yields NC-algorithms for many problems, which are known to be NP-complete in general [Bod89].

Consider the following question. Suppose one is given a circuit (Boolean or arithmetic) of size $s$ and bounded fan-in, for which the underlying graph has bounded treewidth. Does this imply, as intuition might suggest, there must exist an equivalent bounded fan-in circuit of size $poly(s)$ and depth $O(\log s)$ ? We show that in the Boolean case the situation is as expected, which yields the following theorem:

**Theorem 1.1** *The class of languages accepted by non-uniform constant fan-in circuits of polynomial size and bounded treewidth equals non-uniform* $NC^1$.

Due to a celebrated result of Barrington [Bar86], it is known that $NC^1$ can be simulated by constant width branching programs, which are skew circuits of constant width. A con-stant width circuit can be evaluated using $O(1)$ memory, and hence $\mathrm{SC}^0 = \mathrm{NC}^1$ ($\mathrm{SC}^0$ is the class of Boolean functions computable by constant width circuits of *poly* size, c.f. [Mah07]). Theorem 1.1 strengthens this statement in the non-uniform setting.

---

[*]Institute for Theoretical Computer Science, Tsinghua University, Beijing, China. Email: {mjjansen, jayalal}@tsinghua.edu.cn. This work was supported in part by the National Natural Science Foundation of China Grant 60553001, and the National Basic Research Program of China Grant 2007CB807900,2007CB807901.

[1]For a definition see Section 2.

For arithmetic circuits, the short answer is that the equivalent circuit need not exist: a depth $O(\log s)$ circuit of bounded fan-in computes a polynomial of degree $s^{O(1)}$, but using repeated multiplication a bounded treewidth circuit of size $s$ can easily compute a polynomial of degree $2^s$. We rephrase the question to avoid this triviality.

The class of families of polynomials $\{p_n\}_{n \geq 1}$ of polynomial degree with polynomial size arithmetic circuits is known as VP (See e.g. [BCS97]). Let $\mathrm{VP}[tw = O(\log^i n)]$ stand for the class corresponding to polynomial size circuits of treewidth $O(\log^i n)$, and let $\mathrm{VNC}^1$ denote the class corresponding to bounded fan-in arithmetic circuits of depth $O(\log n)$. Our question becomes the following: is $\mathrm{VP}[tw = O(1)] \subseteq \mathrm{VNC}^1$ ?

The above question remains to be wide open. Here, we show that *multiplicatively disjoint*[2] circuits of size $n^{O(1)}$ and bounded treewidth can be simulated by bounded fan-in formulas of depth $O(\log n)$. In our notation this is stated as follows:

**Theorem 1.2** md-$\mathrm{VP}[tw = O(1)] = \mathrm{VNC}^1$.

From this, we derive the analogous statement for *syntactically multilinear* circuits. The resulting formulas will be syntactically multilinear (denoted with the prefix sm-) as well. This implies the lower bounds by Raz [Raz04] hold all the way up to syntactically multilinear bounded treewidth circuits. We prove

**Theorem 1.3** sm-$\mathrm{VP}[tw = O(1)] = \mathrm{sm}\text{-}\mathrm{VNC}^1$.

Above theorems shed some light on issues regarding bounded width circuits. One has the hierarchy of classes $\{\mathrm{VSC}^i\}_{i \geq 0}$, where $\mathrm{VSC}^i$ corresponds to arithmetic circuits of width $O(\log^i n)$ and size $n^{O(1)}$, for $i \geq 0$. We prove the following (and the Boolean analogue):

**Theorem 1.4**

1. $\mathrm{VSC}^0 \subseteq \mathrm{VP}[tw = O(1)] \subseteq \mathrm{VSC}^1$.

2. $\mathrm{VSC}^i \subseteq \mathrm{VP}[tw = O(\log^i n)] \subseteq \mathrm{VSC}^{i+1}[size = n^{O(\log \log n)}]$, *for any* $i \geq 1$.

Arithmetic circuit width is a fundamental notion, but it is still ill-understood in its relation to other resources. Recently, it has gained renewed attention from several researchers. It more or less embodies a notion of space in the arithmetic setting (See [MR09]). Mahajan and Rao [MR08] study the class $\mathrm{VSC}^0[deg = n^{O(1)}]$, which is obtained from $\mathrm{VSC}^0$ by requiring formal degrees of circuits to be $n^{O(1)}$. Arvind, Joglekar, and Srinivasan [AJS09] give lower bounds for *monotone* arithmetic circuits of constant width.

Theorem 1.3 strengthens the main result of Mahajan and Rao [MR08], which states that *poly* size syntactically multilinear circuits of constant width can be simulated by *poly* size circuits of *log* depth (but it was explicitly left open whether the latter could be ensured to be syntactically multilinear). Theorem 1.2 strengthens Theorem 4 in [JR09], which states that md-$\mathrm{VSC}^0 = \mathrm{VNC}^1$.

---

[2]We indicate this with the prefix md-. See Section 2 for a definition.

In [MR08] the fundamental question is raised whether $VSC^0[deg = n^{O(1)}] \subseteq VNC^1$. As was mentioned, they show this holds under restriction of syntactic multilinearity. To make progress, we demonstrate a different restriction under which an efficient simulation by arithmetic $O(\log n)$ depth formulas is achievable. We apply Theorem 1.2 to give the following result for circuits with *bounded iterated multiplication chains* (For a definition see Section 4):

**Theorem 1.5** *Constant width arithmetic circuits of size $n^{O(1)}$ with constant bounded iterated multiplication chains can be simulated by fan-in two arithmetic formulas of depth $O(\log n)$.*

As a final application, we consider the REACHABILITY problem. Given a (directed or undirected) graph $G = (V, E)$ and $s, t \in V$, this problem asks to test if $t$ is reachable from $s$ in $G$. REACHABILITY captures space bounded computation in a natural way. For directed graphs it is complete for NL [Jon75, Sav73]. The case of undirected graphs was settled recently by Reingold [Rei05] by giving a log-space algorithm for the problem. This shows the problem is complete for L. There has been extensive research aimed at settling the complexity of testing reachability on restricted graphs [ADR05, BTV07, KV09, ABC+06, AL98]. See [All07] for more details.

LogCFL and LogDCFL are the classes of languages that are logspace many-one reducible to non-deterministic and deterministic context-free languages, respectively. LogDCFL can be also characterized as the class of languages that can be recognized by a logspace Turing machine that is also provided with a stack, which runs in polynomial time. It follows by definition that $L \subseteq LogDCFL \subseteq LogCFL$ and $L \subseteq NL \subseteq LogCFL$. However, it is unknown how NL and LogDCFL can be compared. In essense, this asks for a trade-off, trading non-determinism with stack access. Given that directed reachability is an NL-complete problem, giving a LogDCFL upper bound achieves such a trade-off for a restricted class of NL-computations. We show the following:

**Theorem 1.6** REACHABILITY *for directed graphs of bounded treewidth is in* LogDCFL, *provided the tree decomposition is given at the input.*

## 2   Preliminaries

We briefly recall basic circuit definitions. A *Boolean circuit* is a directed acyclic graph, with labels $\{0, 1, x_1, \ldots, x_n, \wedge, \vee, \neg\}$ on its nodes. Nodes with label from $\{0, 1, x_1, \ldots, x_n\}$ are called *input gates*, and designated nodes of zero out-degree are called the *output gates*. The *fan-in* of a gate is its in-degree. Formulas are circuits for which the out-degree of each gate is at most one. We use the standard definitions of the size, width, depth and degree of a circuit. Fan-in is assumed to be bounded. The class $NC^1$ is the class of boolean functions on $n$ bits which can be computed by boolean circuits of depth $O(\log n)$ and size $n^{O(1)}$. $SC^i$ denotes the class of functions computed by polynomial size circuits of width $O(\log^i n)$.

For *arithmetic circuits* over a ring $R$, nodes are labeled by ring constants, formal variables from a set $X$, and $\{+, \times\}$. We assume fan-in is bounded by two. The output of an arithmetic circuit is a polynomial in the ring $R[X]$, defined in the obvious way. The size of a circuit is taken to be the number of $\{+, \times\}$-gates. For a circuit $\Phi$ with designated output gate $f$, the polynomial computed by the output gate is denoted with $\lceil \Phi \rceil$. We denote the set of variables using in $\Phi$ by $Var(\Phi)$. Similarly we use $Var(p)$, if $p$ is a polynomial. Note that $Var(\lceil \Phi \rceil) \subseteq Var(\Phi)$. An arithmetic circuit is called *syntactically multilinear*, if for each multiplication gate the subcircuits rooted at its inputs carry disjoint sets of variables. The *formal degree* of a circuit is defined inductively by taking variable and constant labeled gates to be of degree one. For addition gates one takes the maximum of the degrees of its inputs. For multiplication gates one takes the sum of the degrees. The degree of the circuit is taken to be the maximum degree of a gate.

A *p-family of polynomials* is given by $\{f_m\}_{m \geq 1}$, where $f_m \in R[x_1, x_2, \ldots, x_{p(m)}]$, and where $\deg(f_m) \leq q(m)$, for some polynomials $p$ and $q$. Arithmetic circuit classes contain *p*-families. VP and $\text{VP}_e$ are the classes of *p*-families computable by arithmetic circuits and formulas, respectively, of size $n^{O(1)}$ (See e.g. [BCS97]). For $i \geq 0$, $\text{VSC}^i$ is the class of all *p*-families computable by arithmetic circuits of width $O(\log^i n)$ and size $n^{O(1)}$. In [MR08] the class a-sSC$^i$ is considered, which corresponds to width $O(\log^i n)$ circuits of size and formal degree $n^{O(1)}$. We will denote this class by $\text{VSC}^i[deg = n^{O(1)}]$. The class $\text{VNC}^i$ is the collection of all *p*-families computable by arithmetic circuits of depth $O(\log^i n)$ and size $n^{O(1)}$.

Next we define various graph parameters. The *width* of a layered graph is the maximum number of vertices in any particular layer. A *tree decomposition* of a graph $G = (V, E)$ is given by a tuple $(T, (X_d)_{d \in V[T]})$, where $T$ is a tree, each $X_d$ is a subset of $V$ called a *bag*, satisfying 1) $\bigcup_{d \in V[T]} X_d = V$, 2) For each edge $(u, v) \in E$, there exists tree node $d$ with $\{u, v\} \subseteq X_d$, and 3) For each vertex $u \in V$, the set of tree nodes $\{d : u \in X_d\}$ forms a connected subtree of $T$. Equivalently, for any three vertices $t_1, t_2, t_3 \in V[T]$ such that $t_2$ lies in the path from $t_1$ to $t_3$, it holds that $X_{t_1} \cap X_{t_3} \subseteq X_{t_2}$.

The *width* of the tree decomposition is defined as $\max_d |X_d| - 1$. The *treewidth* tw$(G)$ of a graph $G$ is the minimum width of a tree decomposition of $G$. For a rooted tree $T$, let $X_{\leq t} = \cup_{u \in S_t} X_u$, where $S_t = \{u : u = t$ or $t$ is an ancestor of $u\}$.

**Lemma 2.1** *(Theorem 4.3 in [Bod89]) Let $G = (V, E)$ be a graph with $|V| = n$ and treewidth at most $k$. Then $G$ has a tree decomposition $(T, (X_d)_{d \in V[T]})$ of width $3k + 2$ such that $T$ is a binary tree of depth at most $2\lceil \log_{\frac{5}{4}} n \rceil$.*

**Proposition 2.2** *A leveled graph $G$ of width $k$ has a treewidth at most $2k - 1$.*

**Proof.** Let the levels of $G$ be given by the sets of nodes $L_0, L_1, \ldots, L_m$, The required tree $T$ is just a path $t_1, t_2, \ldots, t_m$ such that $\forall j \ \ X_{t_j} = L_{j-1} \cup L_j$. Clearly, for any $v \in V[G]$, the set of $t \in V[T]$ such that $v \in X_t$ form just an edge in $T$, and hence is connected. Every edge in $G$ is covered by definition, and hence the this gives a valid tree decomposition of the graph $G$.

# 3 Arithmetic Circuits of Bounded Treewidth

The treewidth of a circuit with underlying graph $G$ is defined to be tw$(G)$. We introduce the class $\text{VP}[tw = O(\log^i n)]$ as the class of *p*-families of polynomials $\{f_n\}_{n \geq 1}$ that can be computed by fan-in two arithmetic circuits of size $n^{O(1)}$ and treewidth $O(\log^i n)$. As Theorem 1.4 states, these classes interleave with the $\text{VSC}^i$ classes. We postpone the proof of Theorem 1.4 as it uses developments of our main construction.

**Theorem 3.1** *For any multiplicatively disjoint arithmetic circuit $\Phi$ of size $s$ and treewidth $k$, there exists an equivalent formula $\Gamma$ of size at most $s^{O(k^2)}$.*

**Proof.** Let $\Phi$ be a multiplicatively disjoint circuit of size $s$, and let $(T, (X_d)_{d \in V[T]})$ be a tree decomposition of $\Phi$ of width $k$. By Lemma 2.1, we can assume that $T$ is a rooted binary tree of depth $d = O(\log s)$. We first preprocess $T$ and $\Phi$ using Proposition 3.2. For a proof see Appendix A.

**Proposition 3.2** *There exists a circuit $\Phi'$ of size at most $2s$, for which $\lceil \Phi \rceil = \lceil \Phi' \rceil$, with tree decomposition $(T', (X'_d)_{d \in V[T']})$ of width at most $k' = 3k + 2$ and depth at most $d$, so that for any $t \in T'$, for any non-input gate $g \in X'_t$ with inputs $g_1$ and $g_2$, either both $g_1, g_2 \in X'_t$ or both $g_1, g_2 \notin X'_t$. In the latter case it holds that $g_1 \notin X'_{\leq t}$ iff $g_2 \notin X'_{\leq t}$.*

We assume wlog. that $\Phi$ and $(T, (X_d)_{d \in V[T]})$ satisfy the conditions of Proposition 3.2, as the increase in $k$ and $s$ due to preprocessing does not affect the bound we are aiming for. For any tree node $t \in T$ and $f \in X_t$, we define a circuit $\Phi_t$, which is obtained from the subgraph $\Phi[X_{\leq t}]$, by turning all $g \in X_t$ that take both inputs from gates not in $X_{\leq t}$ into input gates with label $z_g$. For any $f \in X_t$, let $\Phi_{t,f}$ be the subcircuit of $\Phi_t$ rooted at gate $f$. At most $k + 1$ new $z$-variables will be used at the tree node $t$. Crucially, observe that, since $\Phi$ is multiplicatively disjoint, any gate in $\Phi_{t,f}$ computes a polynomial that is multilinear in $z$.

We will process the tree decomposition going bottom up. At a node $t$, we want to compute for each $f \in X_t$ a formula $\Gamma_{t,f}$ equivalent to $\Phi_{t,f}$. Wlog. we assume the output gate of $\Phi$ is contained in $X_r$, for the root $r$ of $T$. Hence, when done, we have a formula equivalent to $\Phi$. As will turn out, in order to keep the size of the computed formulas properly bounded, we require a constant bound on the number of appearances of a $z$-variable in $\Gamma_{t,f}$. We achieve this by brute-force with Proposition 3.3, at the cost of blowing up the size by a factor of $2^{k+1}$. To verify its correctness, observe that the lhs. and rhs. are multilinear polynomial in $F[\overline{x}][z_1, z_2, \ldots, z_{k+1}]$ taking identical values on $\{0, 1\}^{k+1}$, and hence must be identical.

**Proposition 3.3** *Let $f(\overline{x}, z_1, z_2, \ldots, z_{k+1})$ be a polynomial that is multilinear in $z$, then $f(\overline{x}, z_1, z_2, \ldots, z_{k+1}) = \sum_{b \in \{0,1\}^{k+1}} \left( \prod_{i \in [k+1]} (1 - z_i)^{1-b_i} z_i^{b_i} \right) f(\overline{x}, b_1, b_2, \ldots, b_{k+1})$.*

The recursive procedure for computing the desired formula equivalent to $\Phi_{t,f}$ is given by Algorithm 1. Formally, for any $t \in T$, and $f \in X_t$, let $\Gamma_{t,f}$ be the formula output by the procedure call $Traceback(t, f)$. The following lemma proves its correctness:

**Lemma 3.4** *For any $t \in T$, and any $f \in X_t$, $\lceil \Gamma_{t,f} \rceil = \lceil \Phi_{t,f} \rceil$.*

**Proof.** The proof will proceed by structural induction both on $T$ and $\Phi$. The statement can be easily verfied for the two base cases: if $t$ is a leaf of $T$, or $f$ is an input gate in $\Phi$. For the induction step, suppose $t$ has children $t_0$ and $t_1$, and say $f = f_0 \circ f_1$, with $\circ \in \{+, \times\}$. We ignore line 17 of the procedure $Traceback$, since it does not modify the output of the computed formula.

In case both $f_0, f_1 \in X_t$, by induction hypothesis, $\lceil \Gamma_{t,f_0} \rceil = \lceil \Phi_{t,f_0} \rceil$ and $\lceil \Gamma_{t,f_1} \rceil = \lceil \Phi_{t,f_1} \rceil$. Observe that in this case $Traceback(t, f)$ returns $\Gamma_{t,f_0} \circ \Gamma_{t,f_1}$, so $\lceil \Gamma_{t,f} \rceil = \lceil \Gamma_{t,f_0} \circ \Gamma_{t,f_1} \rceil = \lceil \Gamma_{t,f_0} \rceil \circ \lceil \Gamma_{t,f_1} \rceil = \lceil \Phi_{t,f_0} \rceil \circ \lceil \Phi_{t,f_1} \rceil = \lceil \Phi_{t,f} \rceil$.

Now assume not both $f_0, f_1 \in X_t$. By Proposition 3.2, this means $f_0 \notin X_t$ and $f_1 \notin X_t$. Furthermore, we either have $f_0, f_1 \in X_{\leq t}$, or $\{f_0, f_1\} \cap X_{\leq t} = \emptyset$. In the latter case, $\lceil \Phi_{t,f} \rceil = z_f$, which is exactly what is returned by $Traceback(t, f)$. In the former case, say $f_0 \in X_{\leq t_{i_1}}$ and $f_1 \in X_{\leq t_{i_2}}$, for $i_1, i_2 \in \{0, 1\}$. Observe that by the tree decomposition properties $f \in X_{t_{i_1}}$, which makes the call of $Traceback(t, f)$ on line 11 valid. Note that $f_0 \notin X_{\leq t_{i_2}}$ and $f_1 \notin X_{\leq t_{i_1}}$. Hence, by the tree decomposition properties, if $i_1 \neq i_2$, there would exist a node $t'$ with $t_1$ as ancestor such that $f, f_0 \in X_{t'}$, but $f_1 \notin X_{t'}$. Due to Proposition 3.2 this case does not arise.

The algorithm first computes $\Gamma = Traceback(t_{i_1}, f)$. By the induction hypothesis $\lceil \Gamma \rceil = \lceil \Phi_{t_{i_1}, f} \rceil$. In $\Phi_{t_{i_1}, f}$, whenever a gate $g$ takes an input from a gate not in $X_{\leq t_{i_1}}$, i.e. by Proposition 3.2 this means both its inputs are not in $X_{\leq t_{i_1}}$, it appears as input node with label $z_g$. However, for the circuit $\Phi_{t, f}$ node $g$ roots $\Phi_{t, g}$. Observe that this means that substituting $\lceil \Phi_{t, g} \rceil$ for each $z_g \in Var(\lceil \Phi_{t_{i_1}, f} \rceil)$ in $\lceil \Phi_{t_{i_1}, f} \rceil$ yields $\lceil \Phi_{t, f} \rceil$. Observe that the tree decomposition properties give us that $g \in X_t$, whenever we make the call on line 13 to compute $\Gamma'$, and hence that this call is valid. By the induction hypothesis, $\lceil \Gamma' \rceil = \lceil \Phi_{t, g} \rceil$. Hence replacing, for all $z_g \in Var(\lceil \Gamma \rceil)$, each gate in $\Gamma$ labeled with $z_g$ by the formula $\Gamma'$ gives a new formula $\Gamma$ satisfying $\lceil \Gamma \rceil = \lceil \Phi_{t, f} \rceil$. □

We must bound the size of the formula $\Gamma_{t, f}$. The idea is that since the number of $z$-variables in formulas is made to depend on $k$ only, one has that $\Gamma_{t, f}$ is constructed from $\alpha(k)$ many copies of $\Gamma_{t_i, g}$ for children $t_i$ of $t$, for some function $\alpha(k)$ depending on $k$ only. This implies a blow-up by a factor of $\alpha(k)$ to go up one level in $T$. In the actual analysis we find $\alpha(k) = 2^{O(k^2)}$. More details can be found in Appendix B.

**Lemma 3.5** *Let $t \in T$ be a node at height $h$, then for any $f \in X_t$, $\Gamma_{t, f}$ has at most $\alpha^h 2^{k+1}$ many gates, where $\alpha = 2^{3k^2 + 9k + 6}$.*

Since $T$ has depth $O(\log s)$, we conclude the final formulas given at the root of $T$ will be of size $s^{O(k^2)}$. □

The proof of Theorem 1.2 is now clear. Trivially $VP_e \subseteq md\text{-}VP[tw = O(1)]$. The converse inclusion follows from Theorem 3.1. Now use the fact that $VP_e = VNC^1$ [Bre74].

## 3.1 Proof of Theorem 1.3

Observe that $sm\text{-}VNC^1 \subseteq sm\text{-}VP_e \subseteq sm\text{-}VP[tw = O(1)]$. For the other direction, let $\Phi$ be a syntactically multilinear circuit of treewidth $k$. We first modify it so that any gate $g$ computing a field constant $\alpha$ is replaced by an input gate $g'$ labeled with $\alpha$. This can be done by removing edges fanning into $g$ and relabeling. Hence the treewidth of the modified circuit is at most $k$. Next, any gate $g$ labeled with a field constant $\alpha$, with edges going to gates $f_1, f_2, \ldots, f_m$, is replaced by $m$ separate copies of $g_1, g_2, \ldots, g_m$, each labeled with $\alpha$, where we add edges $(g_i, f_i)$, for all $i \in [m]$. This does not increase the treewidth, as it can be thought of as a two step procedure, neither of which increases treewidth: first removing the vertex $g$ and attached edges, secondly, adding back the isolated copies. Observe that now we have obtained an equivalent circuit $\Phi'$ that is multiplicatively disjoint. Namely, for purpose of contradiction, suppose there exists a multiplication gate $f = f_1 \times f_2$ such that both $f_1$ and $f_2$ are reachable from some gate $h$. Then there exists such an $h$ for which the paths to $f_1$ and $f_2$ are edge disjoint. For this $h$, since $\Phi'$ is syntactically multilinear, there cannot be variables in the subcircuit $\Phi'_h$. Hence $h$ is a gate computing a constant. Since the paths to $f_1$ and $f_2$ are edge disjoint, $h$ must have out-degree at least two. This contradicts the fact that any gate computing a constant in $\Phi'$ has

**Algorithm 1** Recursive procedure for computing $\Gamma_{t,f}$

---

1: **procedure** $Traceback(t \in T, f \in X_t)$
2: **if** $t$ is a leaf or $f$ is an input gate in $\Phi$ **then**
3:     **return** a formula equivalent to $\Phi_{t,f}$ of size at most $2^{k+1}$ computed by 'brute force'.
4: **else**
5:     let $t_0$ and $t_1$ be the children of $t$ in $T$, and say $f = f_0 \circ f_1$, with $\circ \in \{+, \times\}$.
6:     **if** both $f_0$ and $f_1$ are in $X_t$. **then**
7:         let $\Gamma = Traceback(t, f_0) \circ Traceback(t, f_1)$.
8:     **else**
9:         // Neither $f_0$ nor $f_1$ is in $X_t$, by pre-processing.
10:         If $f_0$ and $f_1$ are not in $X_{\leq t}$ **return** a single node with label $z_f$. Otherwise, say $f_0 \in X_{\leq t_{i_1}}$ and $f_1 \in X_{\leq t_{i_2}}$, for $i_1, i_2 \in \{0,1\}$.
11:         $\Gamma = Traceback(t_{i_1}, f)$.
12:         **for all** $z_g \in Var(\lceil \Gamma \rceil)$ **do**
13:             let $\Gamma' = Traceback(t, g)$.
14:             replace any gate in $\Gamma$ labeled with $z_g$ by the formula $\Gamma'$.
15:         **end for**
16:     **end if**
17:     Process $\Gamma$ to make any $z$-variable occur at most $2^{k+1}$ times using Proposition 3.3.
18:     **return** $\Gamma$.
19: **end if**

---

out degree one. The statement sm-VP$[tw = O(1)] \subseteq$ VNC$_1$ now follows from Theorem 3.1 and the fact that VP$_e$ = VNC$^1$ [Bre74].

To get the strengthened conclusion that sm-VP$[tw = O(1)] \subseteq$ sm-VNC$_1$, we will now indicate how to modify Algorithm 1 to ensure syntactic multilinearity. We use the notation of the proof of Theorem 3.1. Assume we have done preprocessing as indicated above. We know each circuit $\Phi_{t,f}$ is syntactically multilinear, for all $t \in T$, and $f \in X_t$. The goal is to establish inductively that each $\Gamma_{t,f}$ is syntactically multilinear, for all $t \in T$, and $f \in X_t$.

At the base case, i.e. line 3 of Algorithm 1, we can simply enforce the condition by brute force. At line 7, by induction $\Gamma_{t,f_0}$ and $\Gamma_{t,f_1}$ are syntactically multilinear. If $\circ = +$, then so is $\Gamma$. In case $\circ = \times$, whenever the formulas $\Gamma_{t,f_0}$ and $\Gamma_{t,f_1}$ share a variable $\alpha$, since we know $\lceil \Gamma \rceil = \lceil \Phi_{t,f} \rceil$ is multilinear, $\alpha$ does not appear in at least one of the polynomials $\lceil \Gamma_{t,f_0} \rceil$ and $\lceil \Gamma_{t,f_1} \rceil$. Setting $\alpha$ to zero in the corresponding formula ensures $\Gamma$ is syntactically multilinear.

We now argue how to correctly deal with the substitution on line 14, and the processing of $z$ variables on line 17. Consider $\Gamma$ as computed on line 11. We want to ensure it is in the following standard form: $\sum_{a \in \{0,1\}^{k+1}} \left( \prod_{i \in [k+1]} z_i^{a_i} \right) f_a(\overline{x})$, for certain polynomials $f_a \in F[X]$. For this we use the following modification of Proposition 3.3, which is obtained by multiplying out the factors $\prod_{i \in [k+1]} (1 - z_i)^{1-b_i} z_i^{b_i}$. For $a, a' \in \{0,1\}^{k+1}$, we say $a' \leq a$ iff $\{i : a'_i = 1\} \subseteq \{i : a_i = 1\}$. We denote the size of $\{i : a'_i = 1\}$ by $|a'|$.

**Proposition 3.6** *Let* $f(\overline{x}, z_1, z_2, \ldots, z_{k+1})$ *be a polynomial that is multilinear in* $z$. *If we write* $f(\overline{x}, z_1, z_2, \ldots, z_{k+1}) = \sum_{a \in \{0,1\}^{k+1}} \left( \prod_{i \in [k+1]} z_i^{a_i} \right) coef(f, z_1^{a_1} z_2^{a_2} \ldots z_{k+1}^{a_{k+1}})$, *then it holds that* $coef(f, z_1^{a_1} z_2^{a_2} \ldots z_{k+1}^{a_{k+1}}) = \sum_{a' \leq a} (-1)^{|a'|} f(\overline{x}, a')$.

If we use the above proposition to process $z$-variables on line 17, then by induction, $\Gamma$ on line 11 will indeed have the required form, or for simplicity one can also assume we do an extra step of $z$-variable processing. That is, assume we apply above proposition to get $\Gamma$ in the required form. This requires at most $(2^{k+1})^2$ copies of $\Gamma$ and blows up $\Gamma$ by an inconsequential factor of $2^{O(k)}$. Observe that this leaves $\Gamma$ syntactically multilinear.

Now consider line 14. First of all, any $z_g \in Var(\Gamma) \backslash Var(\ulcorner \Gamma \urcorner)$ can be set to zero in $\Gamma$. For the remaining $z$-variables, we claim that for any pair $z_g, z_h \in Var(\ulcorner \Gamma \urcorner)$, whenever $\Gamma_{t,g}$ and $\Gamma_{t,h}$ share a variable $\alpha$, then $coef(\ulcorner \Gamma \urcorner, m) = 0$, for any multilinear monomial in the $z$-variables of $\Gamma$ that contains both $z_g$ and $z_h$. Hence we can remove these terms from the standard form of $\Gamma$, and avoid multilinearity conflicts among products between each of the substituted formulas.

We will verify this claim using the notion of a *proof tree*. A proof tree rooted at a gate $g$ in a circuit $C$, is any tree obtained by recursively selecting gates, starting with $g$, as follows: 1) at an addition gate select exactly one of its children, and 2) at a multiplication gate select both children. We will consider proof trees of $\Phi_{t_{i_1}, f}$ rooted at $f$. For a subset $Z$ of $z$-variables in $\Phi_{t_{i_1}, f}$, we let $\mathrm{PTree}(Z)$ stand for the collection of proof trees rooted at $f$ that have precisely the $z$-variables in $Z$ appearing at its leaves. Given $T \in \mathrm{PTree}(Z)$, let $p(T)$ denote the product of all $X$ variables appearing in $T$. The following proposition is easily proved by structural induction to the circuit $\Phi_{t_{i_1}, f}$.

**Proposition 3.7** *For any multilinear monomial $m$ in $z$-variables used in $\Phi_{t_{i_1}, f}$, it holds that $coef(\ulcorner \Phi_{t_{i_1}, f} \urcorner, m) = \sum_{T \in PTree(Z)} p(T)$, where $Z$ is the set of $z$-variables of $m$.*

Recall that by induction $\ulcorner \Gamma \urcorner = \ulcorner \Phi_{t_{i_1}, f} \urcorner$. Now consider any multilinear monomial $m$ in $z$-variables of $\ulcorner \Phi_{t_{i_1}, f} \urcorner$ with both $z_g$ and $z_h$ in it, where $\Gamma_{t,g}$ and $\Gamma_{t,h}$ share a variable $\alpha$. For purpose of contradiction suppose $coef(\ulcorner \Phi_{t_{i_1}, f} \urcorner, m) \neq 0$. By Proposition 3.7 this means there exists a proof tree in $\Phi_{t_{i_1}, f}$ rooted at $f$ that contains both $z_g$ and $z_h$. This implies $g$ and $h$ are reachable from a single multiplication gate $r$ in $\Phi_{t_{i_1}, f}$, and hence also in $\Phi_{t,f}$. Observe that our construction satisfies the property that for any $t \in V[T]$ and $f \in X_t$, $Var(\Gamma_{t,f}) \subseteq Var(\Phi_{t,f})$. Hence $\alpha$ appears in both $\Phi_{t,g}$ and $\Phi_{t,h}$. Observe that both $\alpha$'s must be reachable from $r$ in $\Phi_{t,f}$. This contradicts the fact that $\Phi_{t,f}$ is syntactically multilinear.

Similarly, one can verify that whenever for a variables $z_g \in Var(\ulcorner \Gamma \urcorner)$, the formula $\Gamma_{t,g}$ contains a variable $\alpha$, then $coef(\ulcorner \Gamma \urcorner, m)$ does not contain $\alpha$ for any monomial $m$ containing $z_g$. Hence any occurrence of $\alpha$ in the formula $\sum_{a' \leq a} (-1)^{|a'|} \Gamma(\overline{x}, a')$ used to compute $coef(\ulcorner \Gamma \urcorner, m)$ can be replaced by zero.

We conclude that under above modifications, Algorithm 1 yields a syntactically multilinear formula $\Gamma_{t,f}$ equivalent to $\Phi_{t,f}$. The proof is completed with the observation of [MR08] that Brent's construction [Bre74], which shows $\mathrm{VP}_e \subseteq \mathrm{VNC}^1$, preserves syntactic multilinearity. □

## 3.2 Proof of Theorem 1.4

The first inclusion of items 1. and 2. follows from Proposition 2.2. For the second inclusion of items 1. and 2, consider a circuit $\Phi$ of size $s$ having tree decomposition $(T, (X_d)_{d \in V[T]})$ of width $k$. By Lemma 2.1, we can assume that $T$ is a rooted binary tree of depth $d = O(\log s)$, and we assume wlog. we have already applied Proposition 3.2 for preprocessing. For $t \in T$ and $f \in X_t$, define $\Phi_{t,f}$ as in the proof of Theorem 3.1. We will argue how to obtain small width circuits $\Psi_{t,f}$ satisfying $\lceil \Psi_{t,f} \rceil = \lceil \Phi_{t,f} \rceil$. As before, we assume wlog. the output gate of $\Phi$ is contained in the bag at the root of $T$, so that when done a small width circuit computing $\lceil \Phi \rceil$ is known.

For any leaf $t \in T$ and $f \in X_t$, we trivially have an equivalent circuit for $\Phi_{t,f}$ of size $k+1$ and width $k+1$.

Now consider $t \in T$ with children $t_0$ and $t_1$. Suppose a set of equivalent circuits $S' = \{\Psi_{t_0,g}\}_{g \in X_{t_0}} \cup \{\Psi_{t_1,g}\}_{g \in X_{t_1}}$ has already been computed for the set of circuits $S = \{\Phi_{t_0,g}\}_{g \in X_{t_0}} \cup \{\Phi_{t_1,g}\}_{g \in X_{t_1}}$.

Let $h_1, h_2, \ldots, h_{k+1}$ be the sequence of nodes in $X_t$ as they appear in an arbitrarily selected topological sort of $\Phi$. Let $w$ be an upper bound on the width of any component in $S'$. For all $i \in [k+1]$, we will build a circuit $\Psi^i$ of width at most $w + i - 1$, that has $i$ many outputs carrying the values $\lceil \Phi_{t,h_j} \rceil$, for all $1 \leq j \leq i$. Wlog. we allow ourselves addition gates of fan-in one, which are used to pass through values.

For $i = 1$, $h_1$ takes both inputs from outside $X_t$. We observe that either $\lceil \Phi_{t,h_1} \rceil$ is a $z$-variable, or some component in $S'$ already computes $\lceil \Phi_{t,h_1} \rceil$. Namely, consider what $Traceback(t, h_1)$ does in order to compute $\lceil \Phi_{t,h_1} \rceil$. In this case, it either outputs a node with a $z$-label, or it first compute $\Gamma = \Gamma_{t_a,h_1}$, for some $a \in \{0,1\}$, but there would be no substitution performed by $Traceback(t, h_1)$, for any $z_g$ in $\Gamma$, since that would mean there exists a gate $g \in X_t$ such that $h_1$ is reachable from $g$. By the correctness of Algorithm 1, $\lceil \Phi_{t_a,h_1} \rceil = \lceil \Gamma_{t_a,h_1} \rceil = \lceil \Gamma \rceil = \lceil \Phi_{t,h_1} \rceil$. So we can take $\Psi^1$ to be either a single input node labeled with a $z$-variables, or take it as some element of $S'$. In both cases the width is at most $w$.

Inductively, for $i > 1$, suppose we have build the circuit $\Psi_{i-1}$. It has outputs computing $\lceil \Phi_{t,h_j} \rceil$ for $1 \leq j \leq i - 1$. First consider the case that $h_i$ takes both inputs from within $X_t$. Similarly by inspection of what Algorithm 1 does for this case, it can be observed that $\Psi_{t,h_i}$, can be build by adding or multiplying two components $\Psi_{t,h_p}$ and $\Psi_{t,h_q}$ with $p, q < i$. Hence we obtain $\Psi^i$ by adding a single gate to $\Psi^{i-1}$ and passing through all other outputs. Hence, in this case we have $width(\Psi^i) \leq \max(width(\Psi^{i-1}), 1 + i - 1) \leq w + i - 1$

Finally, suppose $h_i$ takes an input not in $X_t$. We then know neither inputs are in $X_t$, since we assume we have applied Proposition 3.2. Let us consider Algorithm 1, to see what $Traceback(t, h_i)$ does to compute $\lceil \Phi_{t,h_i} \rceil$. Either a $z$-variable is returned, which will be easy to handle. Otherwise, it computes $\Gamma$ with $\lceil \Gamma \rceil = \lceil \Phi' \rceil$, for some $\Phi' \in S$. Then for $z$-variables in $\lceil \Gamma \rceil$, values are substituted of form $\lceil \Phi_{t,g} \rceil$, for $g \in X_t$. Observe that $h_i$ is reachable from any such $g \in X_t$ in $\Phi$, and hence $g = h_j$ for some $j < i$.

From this we can conclude that the required $\Psi^i$ either is a $z$-variable, or can be build from one component $\overline{\Psi}$ from $S'$ and feeding outputs of $\Psi^j$ for $j < i$ into $z$-variable gates appearing in $\overline{\Psi}$. We do this by adding a copy of $\overline{\Psi}$ below $\Psi^{i-1}$ and passing alongside $\overline{\Psi}$ the output values of $\Psi^{i-1}$. This makes these values available so we can do the appropriate substitutions of $z$-variables in $\overline{\Psi}$. Observe that this way $\Psi^i$ has width at most $\max(width(\Psi^{i-1}), width(\overline{\Psi}) + i - 1) \leq w + i - 1$.

From the above, we observe that any $\Psi^i$ will be at most a factor $O(k)$ larger than any component in $S'$. We conclude that the width increases additively by $O(k)$, and the size multiplicatively by $O(k)$, in order to go up one level in $T$. We conclude that a width $O(k \log s)$ and size $s^{O(\log k)}$ bound holds for any $\Psi_{r,f}$, with $r$ being the root of $T$ and $f \in X_r$. $\qquad\square$

### 3.3 Evaluation over a Finite Field and Boolean Implications

The observation is that Algorithm 1, when applied over $GF(2)$ to an arbitrary $n$-input arithmetic circuit $\Phi$, will result in a formula $\Gamma$ such that for any $a \in GF(2)^n$, $\lceil \Phi \rceil (a) = \lceil \Gamma \rceil (a)$. For this, no assumptions regarding multiplicative disjointness of $\Phi$ is needed. One can prove this condition using structural induction similarly as in Lemma 3.4. For the processing of the $z$-variables on line 17, observe that we have the following adaption of Proposition 3.3:

**Proposition 3.8** *Let* $f(x_1, x_2, \ldots, x_n, z_1, z_2, \ldots, z_{k+1})$ *be a polynomial over* $GF(2)$. *Define the polynomial* $g = \sum_{b \in \{0,1\}^{k+1}} \left( \prod_{i \in [k+1]} (1 - z_i)^{1 - b_i} z_i^{b_i} \right) f(x_1, x_2, \ldots, x_n, b_1, b_2, \ldots, b_{k+1})$. *Then for any* $a \in GF(2)^{n+k+1}$, $f(a) = g(a)$.

One can generalize this to an arbitrary finite field $F$ of size $q = |F|$, by similarly using brute force on line 17 of Algorithm 1 to make sure any $z$-variables appears at most $q^{k+1}$ times in $\Gamma$. Consequently, we have the following theorem:

**Theorem 3.9** *Let* $F$ *be a finite field, and let* $q = |F|$. *For any arithmetic circuit* $\Phi$ *over* $F$ *of size* $s$ *and treewidth* $k$, *there exists formula* $\Gamma$ *over* $F$ *of size at most* $s^{O(k^2 \log q)}$ *such that* $\Phi$ *and* $\Gamma$ *evaluate to identical values for all inputs from* $F$.

For a proof of the following proposition see Appendix C.

**Proposition 3.10** *For every Boolean circuit* $C$ *of bounded fanin and treewidth* $k$, *there is an arithmetic circuit* $C'$ *over* $GF(2)$ *of treewidth* $3k$ *such that* $\forall x \in \{0,1\}^n$, $C(x) = 1$ *if and only if* $C'(x) = 1$.

We will now derive Theorem 1.1. Given a Boolean circuit of size $s$ and treewidth $k$, first convert it into an arithmetic circuit over $GF(2)$ using Proposition 3.10. Now apply Theorem 3.9 to obtain an arithmetic formula $\Gamma$ over $GF(2)$ of size $s^{O(k^2)}$. Balance this formula down to depth $O(k^2 \log s)$ using [Bre74]. Now do the reverse construction of arithmetization and code out an $\{\wedge, \vee, \neg\}$-formula computing the same function. The final circuit has depth $O(k^2 \log s)$. Thus we have proven Theorem 1.1.

We can use a similar reduction to derive a Boolean analogue of Theorem 1.4. The proof is contained in Appendix D. Let $\mathrm{TWC}^i$ denote the class of Boolean functions computed by Boolean circuits of treewidth $O(\log^i n)$.

**Theorem 3.11** *The following inclusions hold in the non-uniform setting*

1. $\mathrm{SC}^0 \subseteq \mathrm{TWC}^0 \subseteq \mathrm{SC}^1$.

2. $\mathrm{SC}^i \subseteq \mathrm{TWC}^i \subseteq \mathrm{SC}^{i+1}[size = n^{O(\log \log n)}]$, *for all* $i \geq 1$.

## 4 Constant Width Circuits

**Definition 4.1** *An* iterated multiplication chain *of length* $m$ *in a circuit* $\Phi$ *is given by a sequence of gates* $g_0, g_1, \ldots, g_m$, *where all are multiplication gates, except possibly* $g_0$, *such that* *both inputs of $g_i$ are reachable from $g_{i-1}$, for all $i \in [m]$. We denote the length of a longest iterated multiplication chain in $\Phi$ by $\mathcal{M}(\Phi)$.*

Note that if $\mathcal{M}(\Phi) = 0$, then $\Phi$ is multiplicatively disjoint.

**Lemma 4.2** *For any leveled arithmetic circuit $\Phi$ of size $s$, width $w$, with fanout of every gate bounded by two, there exists equivalent multiplicatively disjoint circuit $\Phi'$ of size at most $\frac{s^{\mathcal{M}(\Phi)+2}-1}{s-1} - 1$. The circuit $\Phi'$ has a tree decomposition $(T, (X_t)_{t \in V[T]})$ of width at most $2w-1$, such that for any set $L$ of nodes in $\Phi$ on the same level, there exists $t \in T$ with $L \subseteq X_t$.*

**Proof.** The lemma is proved by induction on $d$. Let the levels of $\Phi$ be given by sets of gates $L_0, L_1, \ldots, L_m$, with edges going from $L_i$ to $L_{i+1}$ only, for $i = 0, 1, \ldots, m-1$. The circuit $\Phi$ has a tree decomposition $(T, (X_t)_{t \in V[T]})$ of width $2w-1$, where $T$ is given by the path $t_1, t_2, \ldots, t_m$, and $X_{t_j} = L_{j-1} \cup L_j$, for $j \in [m]$.

In case $d = 0$, we already have that $\Phi$ is multiplicatively disjoint, and we can conclude the lemma holds. For the induction case, suppose $d > 0$.

We proceed by going from level $r = m-1$ down to $r = 0$. At level $r$, let $(g^i)_{i \in I}$ be the set of gates in $L_r$, such that for every $i \in I$, $g^i \in L_r$ is used by two gates $g_1^i, g_2^i \in L_{r+1}$ such that from $g_1^i$ and $g_2^i$ there is an identical multiplication gate reachable.

If $I$ is non-empty, do the following: Create a copy $\Phi_I$ of the subcircuit of $\Phi$ that consists of the gates $(g^i)_{i \in I}$, together with all the gates these depend on. Note that $\mathcal{M}(\Phi_I) \leq d-1$. Inductively apply the lemma to make $\Phi_I$ multiplicatively disjoint. Modify $\Phi$, by removing the edges $(g^i, g_2^i)$, and adding the edge from $g^i$ in the copy of $\Phi_I$ to $g_2^i$, for all $i \in I$.

Once all levels of $\Phi$ have been processed in the above manner, it is clear the resulting circuit is multiplicatively disjoint. Now we bound the size of this circuit. By induction hypothesis, the copy of $\Phi_I$ has size at most $\frac{s^{d+1}-1}{s-1} - 1$. In the worst case we create such a copy at each level, i.e. at most $s$ many times. Hence the created circuit has size at most $s + s \cdot (\frac{s^{d+1}-1}{s-1} - 1) = \frac{s^{d+2}-1}{s-1} - 1$.

Finally, we verify that the constructed circuit has treewidth at most $2w-1$. We will check that the duplication performed at level $r$ results in a new circuit with a tree decomposition satisfying all the required properties. By induction hypothesis, for the multiplicatively disjoint version of $\Phi_I$ we have tree decomposition $(T', (X_t)_{t \in V[T']})$ of width at most $2w-1$. Furthermore, there exists $t' \in T'$ such that for all $i \in I$, the copy of $g^i$ is in $X_{t'}$. We construct a tree decomposition $(T'', (X_t)_{t \in V[T'']})$ from $T$ and $T'$ as follows. First, create a new tree node $t''$ with bag $X_{t''}$ containing the copy of $g^i$ and $g_2^i$, for all $i \in I$. Next, let $T''$ be the tree obtained by connecting tree node $t_{r+1} \in T$ with $t''$, and connecting $t''$ to $t' \in T'$. The bag $t''$ covers the edges from the copy of $g^i$ to $g_2^i$, for all $i \in I$. Other edges are either covered in $T$ or in $T''$. The decomposition properties can now easily been seen to hold. Note that for any level $L_r$ in $\Phi$ there exists a bag in $T''$ containing $L_r$. Bag sizes in $T''$ are bounded by $2w-1$. ☐

**Theorem 4.3** *For any leveled arithmetic circuit $\Phi$ of size $s$ and width $w$, there exists equivalent formula $\Gamma$ of depth $d = O(w^4 \mathcal{M}(\Phi) \log s)$ and consequently size at most $2^d$.*

**Proof.** Consider any leveled arithmetic circuit $\Phi$ of size $s$ and width $w$. The fanout of every gate in $\Phi$ is bounded by $w$. We first modify $\Phi$ to have the fanout of every gate bounded by two.

This is done by adding dummy 'add to zero' addition gates arranged in complete binary trees of depth at most $\lceil \log w \rceil$. The width of the resulting circuit will be $O(w^2)$, and its size $O(ws)$. Next we apply Lemma 4.2 to obtain an equivalent circuit of size $s^{O(\mathcal{M}(\Phi))}$ and treewidth $O(w^2)$ that is multiplicatively disjoint. Now apply Theorem 3.1 to get an equivalent formula of size $s^{O(w^4 \mathcal{M}(\Phi))}$. Finally, balance this formula using Brent's construction [Bre74] down to depth $O(w^4 \mathcal{M}(\Phi) \log s)$. $\square$

Theorem 1.5 immediate follows from Theorem 4.3. Note that conversely one has the inclusion $\mathrm{VNC}^1 \subseteq \mathrm{VSC}^0[\mathcal{M} = O(1)]$, due to [BC88]. We remark that Theorem 4.3, to the best of our knowledge, when applied to an arbitrary circuit family $\{\Phi_n\}_{n \geq 1}$ of size $n^{O(1)}$, constant width, and $\mathcal{M}(\Phi_n) = o(\log n)$, the resulting $n^{o(\log n)}$ size formulas $\{\Gamma_n\}_{n \geq 1}$ provide the best-known upper bound for the size of equivalent formulas. For $\mathcal{M}(\Phi_n) = O(\log n)$, one is dealing with the general case of the class $\mathrm{VSC}^0[deg = n^{O(1)}]$. In this case our construction yields equivalent $O(\log^2 n)$-depth formulas of size $n^{O(\log n)}$. Formulas with such parameters can also be obtained using the construction in [VSBR83].

# 5 Testing Reachability in Bounded Treewidth Graphs

The following proposition is proved in Appendix E:

**Proposition 5.1** *Given a directed graph $G = (V, E)$ of bounded treewidth and two vertices $s, t \in V$, we can obtain a circuit $C$ of bounded treewidth and an input $x$ such that $C(x) = 1$ if and only if $t$ is reachable from $s$ in the graph $G$.*

For the proof of Theorem 1.6, we compose the recoding of the input of Proposition 5.1 and Proposition 3.10, to get an arithmetic circuit $C$ over $GF(2)$ and $x \in \{0,1\}^{n^2}$, together with its tree decomposition, such that $t$ is reachable from $s$ in $G$ iff $C(x) = 1$. This simple recoding of the input can be computed within *logspace*. Hence the theorem follows by the following observation:

**Proposition 5.2** *Given an arithmetic circuit $C$ over $GF(2)$, its tree decomposition $(T, (X_d)_{d \in V[T]})$ of constant width $k$, and an input $x \in \{0,1\}^n$, testing whether $C(x) = 1$ can be done in* LogDCFL.

*Proofsketch.* The proof proceeds by analyzing the algorithm $Traceback$. We are given the circuit $C$ and an input $x$. We replace each gate of $C$ labeled by $x_i$ with its Boolean value given as input. Next we run $Traceback$ to compute an equivalent formula. A straigtforward analysis gives that this computation takes time polynomial in the length of the input. We claim that in addition we can implement the algorithm using only $O(\log n)$ workspace, provided we use a stack (whose space usage is not counted towards the space bound).

We implement the recursion by using the stack in the usual way. At any point in the recursion, the configuration can be represented with $O(\log n)$ space. Namely, by considering the source-code of $Traceback$, any of the variables, with the exception of $\Gamma, \Gamma'$, and $\Gamma_{t,f}$, are pointers into either the graph or the tree decomposition and hence take $O(\log n)$ size.

The observation is that for any of the equivalent formulas being held by $\Gamma, \Gamma'$, and $\Gamma_{t,f}$, there are no $x$-variables appearing, since these are replaced by Boolean values. $Traceback$

ensures these formulas are represented as multilinear polynomials in $z$-variables over $GF(2)$. For any such formula there are at most $k+1$ different $z_g$'s, since whenever $z_g$ is a variables in $\Gamma_{t,f}$, we always have $g \in X_t$. Furthermore, each $z_g$ appears at most $2^{k+1}$ times. Each $z_g$ has associated a pointer to a node $g$ in the graph. We conclude it takes $O(k2^{k+1}\log n)$ to represent any of $\Gamma, \Gamma'$, and $\Gamma_{t,f}$. This makes it possible to execute the algorithm in polynomial time on a Turing machine with a stack and $O(\log n)$ workspace. The final equivalent formula will simply be the value $C(x) \in \{0, 1\}$. $\qquad\square$

# A    Proof of Proposition 3.2

Let us show how to enforce the first condition. Suppose at some $t \in T$ we have a node $g \in X_t$ with inputs $g_1$ and $g_2$ with $g_1 \in X_t$, but $g_2 \notin X_t$. This is resolved by adding an addition gate $a$ and an input gate $b_{g_2}$ labeled by zero to $\Phi$, removing the edge $(g_2, g)$, and adding edges from $b$ and $g_2$ to $a$, and from $a$ to $g$. $T$ is modified by adding $a$ to every bag containing $g$, and $b$ is added to every bag containing $g_2$. Repeat this procedure until all conflicts are resolved, where for any node $g_2$ we reuse the node $z_{g_2}$ with label zero, if it has already been introduced at a previous operation. Hence for any node $g$ we add at most two addition gates and at most one gate $b_g$ to all bags containing $g$. Hence the width of the new tree decomposition will be at most $3k + 2$. The depth will remain to be $d$. The second condition follows from the first. Namely, suppose $g_1 \in X'_{\leq t} \backslash X'_t$. By the tree decomposition properties it must be that $g$ and $g_1$ are contained in $X'_{t'}$ for some descendant $t'$ of $t$. Hence $g_2 \in X'_{t'}$.    □

# B  Proof of Lemma 3.5

**Proof.** We use $B(h)$ to denote a bound on the size of any $\Gamma_{t,f}$, for a $t$ at height $h$ and $f \in X_t$. If $t$ is a leaf, then we can give a formula for $\Gamma_{t,f}$ of size at most $2^{k+1}$, so we take $B(0) = 2^{k+1}$. Now suppose $height(t) > 0$. Consider unfolding the recursive calls of $Traceback(t, f)$, where we do not unfold calls of $Traceback(t', f)$, for children $t'$ of $t$. Rather, for these we take the upper estimate that there we obtain a formula $\Gamma$ of size at most $B(h-1)$ with at most $k + 1$ many $z$-variables, each of which appears at most $2^{k+1}$ many times in $\Gamma$. We can partition the calls in stages: $Traceback(t, f)$ is at stage 0. Any recursive call made of $Traceback$ at stage $i$ remains at stage $i$, with the exception of calls of $Traceback$ on line 14. These calls signal the next stage, i.e. are at stage $i + 1$. In total the number of stages is bounded by $k + 1$.

For the first stage in the worst case first line 7 is executed up to recursion depth $k+1$, before a call is made for a child of $t$. This gives a formula $\mathcal{F}$ of size at most $(2^{k+1}-1)+2^{k+1}B(h-1) \leq 2^{k+2}B(h-1)$. It has at most $2^{k+1} \cdot (k+1)2^{k+1} \leq 2^{3k+2}$ many occurrences of $z$ variables. For each of the $z$ variables in $\mathcal{F}$ we start unfolding second stage calls of $Traceback$. For each such call we have the same bounds $2^{k+2}B(h-1)$ on the number of tree nodes, and $2^{3k+2}$ for the number of $z$-variables it produces. This process continues up to stage at most $k + 1$. Since the $z$ variables cause a branching of degree at most $2^{3k+2}$, we can state that $B(h) = (2^{k+2}B(h-1)) \cdot 2^{(3k+2)(k+2)} = (2^{3k^2+9k+6})B(h-1)$. This proves the lemma. $\qquad\blacksquare$

# C  Proof of Proposition 3.10

The standard way to obtain $C'$ is by replacing in $C$ each $\wedge(f_1, f_2)$ by $f_1 \times f_2$, each $\neg(f_1)$ by $1 + f_1$, and each $\vee_a(f_1, f_2)$ by $(f_1 +_a f_2) +_b (f_1 \times_c f_2)$. In the latter, we labeled the gates to indicate which vertices are reused. We introduce new vertices and edges only in the case of $\neg$ and $\vee$. Modify the tree decomposition of $C$, to get one for $C'$, as follows: In the case of $\neg$, it suffices to add the new isolated vertex labeled by 1 to one bag containing the $\neg$ gate. In the $\vee$-case it suffices to add newly created vertices $b$ and $c$ to all bags containing $a$. $\qquad\blacksquare$

# D  Proof of Theorem 3.11

The lower bound in (1) and (2) directly follows from Proposition 2.2. To argue the upper bounds, we convert the given TWC$^i$-circuit into an arithmetic circuit $\Phi$ over $GF(2)$ of bounded treewidth using Proposition 3.10. Now applying the construction from the proof of Theorem 1.4 to $\Phi$, will give us a circuit over $GF(2)$ of width $O(\log^{i+1} n)$, and size $n^c$, where $c = O(1)$ if $i = 0$, and $c = n^{O(\log \log n)}$ otherwise. Now construct the required Boolean SC$^{i+1}$-circuit by replacing each gate by a Boolean gadget computing the arithmetic operation over $GF(2)$. $\qquad\blacksquare$

# E  Proof of Proposition 5.1

We describe the standard construction and then argue that it preserves treewidth. Given the graph $G$, the circuit $C$ is obtained by placing an $\vee$ gate at every node $v \in V$. The root gate is the $\vee$ gate placed at $t$. Now for every edge $(u, v) \in E$ place an $\wedge_{uv}$ of fanin 2 and fanout 1 which receives an input from the $\vee_u$ and feeds into $\vee_v$. The second input of $\wedge_{uv}$ is the variable $x_{uv}$ which is essentially the $(u,v)$ entry of the adjacency matrix of $G$ given at the input. The $\vee$ gate placed at $s$ will have an additional input which is assigned to a value 1. It is clear from the construction that this 1 propagates to the output gate if and only if there is a directed path from the node $s$ to the node $t$.

Now we need to argue that this construction preserves treewidth. We will show this constructively. Suppose we are given a tree decomposition $(T, (X_d)_{d \in V[T]})$, such that $\forall t \ |X_t| \leq k$. We will show that the circuit $C$ obtained above also has a tree decomposition with $|X_t| \leq k+2$. The construction simply adds the vertex $\wedge_{uv}$ and the input node $x_{uv}$ to the bag which contains the vertices correpsonding to $\vee_u$ and $\vee_v$. This covers all the new edges introduced, namely $(\vee_u, \wedge_{uv}), (\wedge_{uv}, \vee_v), (x_{uv}, \wedge_{uv})$. The width has increased just by 2. $\qquad\square$